\documentclass[10pt,pra,aps,twocolumn,groupedaddress,showpacs]{revtex4-1}
\usepackage{graphicx}
\usepackage{bm}
\usepackage{amssymb,amsfonts,amsmath}

\begin{document}
\renewcommand{\vec}[1]{\boldsymbol{#1}}
\newcommand{\qq}{\vec{q}}
\newcommand{\xx}{\vec{x}}
\newcommand{\rr}{\vec{r}}
\newcommand{\ud}{\mathrm{d}}
\newcommand{\ui}{\mathrm{i}}
\newcommand{\ue}{\mathrm{e}}
\newcommand{\qx}{q_x}
\newcommand{\qy}{q_y}
\newcommand{\qz}{q_z}
\newcommand{\aad}{a/a_d}
\newcommand{\apf}{a_\text{pb}}
\newcommand{\acrit}{a_\text{crit}}
\newcommand{\RD}{\mathcal{R}_D}
\newcommand{\abs}[1]{\left|#1\right|}
\newcommand{\etal}{\emph{et al.}}
\newcommand{\kB}{k_\mathrm{B}}
\newcommand{\Tc}{T_\text{c}}
\newcommand{\SI}{^\text{(SI)}}

\title{Symmetry-breaking thermally induced collapse of dipolar Bose-Einstein condensates}

\author{Andrej Junginger}
\email{andrej.junginger@itp1.uni-stuttgart.de}
\author{J\"org Main}
\author{G\"unter Wunner}
\affiliation{Institut f\"{u}r Theoretische Physik 1, Universit\"{a}t Stuttgart, 70550 Stuttgart, Germany}

\author{Thomas Bartsch}
\email{T.Bartsch@lboro.ac.uk}
\affiliation{Department of Mathematical Sciences, Loughborough University, Loughborough LE11 3TU, UK}

\date{\today}

\begin{abstract}
We investigate a Bose-Einstein condensate with additional long-range dipolar
interaction in a cylindrically symmetric trap within a variational framework.
Compared to the ground state of this system, little attention has as yet been
payed to its unstable excited states. For thermal excitations, however, the
latter is of great interest, because it forms the ``activated complex'' that
mediates the collapse of the condensate. For a certain value of the s-wave
scatting length our investigations reveal a bifurcation in the transition state,
leading to the emergence of two additional and symmetry-breaking excited states.
Because these are of lower energy than their symmetric counterpart, we predict
the occurrence of a symmetry-breaking thermally induced collapse of dipolar
condensates. We show that its occurrence crucially depends on the trap geometry
and calculate the thermal decay rates of the system within leading order
transition state theory with the help of a uniform rate formula near the rank-2
saddle 
which allows to smoothly pass the bifurcation.
\end{abstract}



\maketitle

\section{Introduction}

Since the first experimental realization of a Bose-Einstein condensate (BEC) in
1995 \cite{Anderson1995}, the field of ultra-cold quantum gases has developed
rapidly. An important milestone in this development was the condensation of
$^{52}$Cr and $^{164}$Dy atoms \cite{Griesmaier2005,Lu2011}, which, due to
their large magnetic dipole moments, interact via the anisotropic, long-range
dipole-dipole interaction (DDI). Because the latter can be either attractive or
repulsive, depending on the  orientation of the dipoles, a wealth of new
phenomena emerges in these BECs, such as stability diagrams that crucially
depend on the trap geometry \cite{Koch2008,Santos2000,Goral2002}, isotropic as
well as anisotropic solitons \cite{Pedri2005,Nath2009,Tikhonenkov2008},
biconcave or structured ground state density distributions
\cite{Dutta2007,Ronen2007, Goral2000}, radial and angular rotons
\cite{Ronen2007,Santos2003,Wilson2008}, as well as anisotropic collapse
dynamics \cite{Metz2009, Lahaye2008}. Investigations of the 
physics of dipolar systems may in the future be extended with the help of
heteronuclear molecules \cite{Kerman2004,Deiglmayr2008,Ni2008,Ospelkaus2008}
or by laser-induced electric dipole-dipole interaction \cite{ODell2003}.

The stability of a dipolar BEC is in general determined by the interplay of
the two-particle interactions present, namely the contact interaction (described
by the s-wave scattering length) as well as the DDI, and the geometry and the
strength of the trap. In the case of an attractive scattering
interaction, the ground state of a harmonically trapped dipolar quantum gas,
which we consider in this paper, is metastable and the BEC can decay by a
coherent collapse of the condensate. The collapse can be induced by macroscopic
quantum
tunnelling at $T=0$ \cite{Stoof1997} or by decreasing the s-wave scattering
length into a region where the BEC cannot exist anymore \cite{Ronen2007}.

Another possibility investigated in this paper is the coherent
collapse
due to thermal excitations of the condensate at finite temperature.
We consider temperatures which are, on the one hand,
small compared to
the critical temperature $\Tc$ where the ground state is populated
macroscopically so that we have an almost pure condensate.
Although modifications will be
caused by the interaction of the bosons, a rough estimate of this regime can
be obtained from the ideal Bose gas in a harmonic trap for which the fraction
$N_0/N$ of condensed particles is given by $N_0/N = 1 - (T/\Tc)^3$
\cite{Pethick2008}. For temperatures $0 < T \lesssim 0.2\,\Tc$ we then have more
than 99\% of the bosons in the condensate and can neglect the influence of the
thermal cloud. For a $^{52}$Cr condensate that we investigate in the
following the critical temperature is $\Tc \approx 700\,$nK
\cite{Griesmaier2005}. We will therefore consider temperatures of $T \lesssim
140\,$nK where the thermal excitations are of collective nature and describe the
quasi-particle modes of the whole condensate.

On the other hand, the
temperature must be high enough so that collective
oscillations of the BEC are activated. As will be discussed below, in the
relevant
region of the scattering length and for experimentally accessible particle
numbers, the frequencies of the collective modes
can be assigned to a temperature on the order of $T \sim 1\,$nK. Thus,
in the temperature regime of several tens of nK the latter are sufficiently
activated.

Note that, at higher temperatures than discussed above, a
significant number of bosons
will
occupy excited states so that the Gross-Pitaevskii equation (GPE) will no more
be adequate to such a
system. In this case Hartree-Fock-Bogoliubov theory
\cite{Proukakis1996,Griffin1996} can be applied, allowing the investigation of
thermally excited BECs at finite temperatures up to the critical temperature.
Note further that, at sub-nK temperatures, where collective
oscillations are not present anymore, macroscopic quantum tunnelling will be
the dominant decay mechanism. Both these temperature regimes are, however, not
subject of this paper.
 
In the temperature regime described above dipolar quantum gases can be well
described by a nonlocal GPE,  
which is usually solved  either numerically or by variational approaches.  The
GPE possesses, apart from the stable ground state, also one or several excited
stationary solutions. To date these solutions have received little attention in
the literature. However, it is exactly these excited states which form the
transition states (TS) on the way to the thermally induced  collapse of the BEC,
and they therefore play a key role in thermally excited condensates.

In this paper we investigate dipolar BECs using a Gaussian variational approach
and reveal a remarkable bifurcation of 
the TS. The physical interpretation of the emerging additional states directly
implies that there exist regions of the physical parameters of the system, i.e.\
the trap frequencies and the s-wave scattering length, in which a
symmetry-breaking thermally induced collapse of the condensate would be
observable in an experiment. 

The BEC's thermal decay rate can be obtained by applying transition state theory
(TST). However, the standard TST rate 
formula fails near bifurcations. With the help of a suitable normal form of the
potential which describes the entire configuration of several saddle points we
will derive a uniform rate formula which solves this problem.

The paper is organized as follows: In Sec.\ \ref{sec-theory-a}, we provide
the description of the dipolar quantum gas
within the variational framework, introduce the equivalent Hamiltonian picture
and discuss the behavior of the potential when one varies the s-wave
scattering length. Sec.\ \ref{sec-theory-b} demonstates the calculation of
the BEC's decay rate, and in Sec.\ \ref{sec-results} we present and discuss the
results.

\section{Theory}
\label{sec-theory}

\subsection{Description of the BEC}
\label{sec-theory-a}

Assuming all dipoles to be aligned along the $z$-direction, we can write the
extended GPE of dipolar BECs in axisymmetric harmonic traps in the form
\begin{align}%
	\ui \partial_{\tilde{t}} \tilde{\psi}(\tilde{\rr},\tilde{t}) &=
	\Bigl( - \Delta + 8 \pi \frac{a}{a_d}
\abs{\tilde{\psi}(\tilde{\rr},\tilde{t})}^2 \nonumber
	+  N^4\gamma_\rho^2 \rho^2 + N^4\gamma_z^2 z^2  \Bigr.
	 \nonumber\\
	& + \Bigl.  \int \! \ud^3 \tilde{\rr}'\, \frac{1-3\cos^2
\theta}{\abs{\tilde{\rr}-\tilde{\rr}'}^3} \,
\abs{\tilde{\psi}(\tilde{\rr}',\tilde{t})}^2 \Bigr) \,
\tilde{\psi}(\tilde{\rr},\tilde{t}) \, .
	\label{eq-GPE}
\end{align}
Here, $\tilde{\psi}(\tilde{\rr},\tilde{t})$ is the scaled condensate wave
function, $\gamma_{\rho,z}= \omega_{\rho,z}/(2\omega_d)$ are the dimensionless
trap frequencies in radial and $z$-direction, $a/a_d$ 
denotes the scaled s-wave scattering length, and $\theta$ is the angle between
the $z$-axis and the vector $\rr - \rr'$. We use
``natural units'' \cite{Koeberle2009a} for the length $a_d = m \mu_0
\mu^2/(2\pi\hbar^2)$, energy $E_d=\hbar^2 / (2 m a_d^2)$, frequency $\omega_d =
E_d/\hbar$ which are defined using the mass $m$ of the bosons, their
magnetic moment $\mu$ and the vacuum permeability $\mu_0$. Furthermore, we
apply a particle number scaling $\rr = N
\tilde{\rr}$, $\psi = N^{-3/2} \tilde{\psi}$, $E = N^{-1} \tilde{E}$, $\beta = N
\tilde{\beta}$, $\omega = N^{-2} \tilde{\omega}$ in order to eliminate the
explicit occurrence of the particle number $N$
in the interaction terms in Eq.\ \eqref{eq-GPE}. Also, in what follows the
inverse temperature is measured by the dimensionless quantity
$\beta=E_d/\kB T$.

Since the unstable excited eigenstates of Eq.\ \eqref{eq-GPE} are not
accessible via imaginary time evolution on a grid, we will resort to a 
variational approach. Because the bosons are trapped harmonically a natural
choice is a Gaussian trial wave function \cite{Huepe1999,Huepe2003,
PerezGarcia1996, PerezGarcia1997, Yi2000, Yi2001, Parker2009, Muruganandam2012}
\begin{equation}
	\tilde{\psi}(\tilde{\rr},\tilde{t}) = \mathcal{N} \prod_{\sigma=x,y,z}
\exp\left(
-\frac{\sigma^2}{8q_\sigma^2} + \ui \frac{p_\sigma \sigma^2}{4 q_\sigma} 
\right)\,,
	\label{eq-ansatz}
\end{equation}
which well approximates the true wave function as long as the interactions
between the bosons are not too strong. Here $\mathcal{N}$ is the normalization
factor of the wave function, $\int
\ud^3 \tilde{\rr}\, |\tilde{\psi}(\tilde{\rr},\tilde{t})|^2 = 1$, and 
$q_\sigma, p_\sigma$ are time-dependent variational functions. Note that the
Cartesian geometry of the ansatz is capable of describing $m=0$ (breathing mode)
and $m=2$ (quadrupole mode) collective oscillations of the condensate, and
therefore covers the two most important modes of the system.

Even though it is well known that the simple ansatz \eqref{eq-ansatz} with a
\emph{single} Gaussian will only yield 
qualitative results, it is crucial because it is the only access to dipolar BECs
that can \emph{globally} be mapped to an equivalent Hamiltonian system $H =
\vec{p}^2/2 + V(\qq)$ \cite{Eichler2011}. The existence of a Hamiltonian,
however, is essential for the application of TST and the derivation of the
subsequent rate formula near a rank-2 saddle, since both are formulated in phase
space. As shown in Ref.\ \cite{Eichler2011} the potential $V(\qq)$ reads
\begin{align}
	V(\qq) = 	&\frac{1}{8\qx} + \frac{1}{8\qy} + \frac{1}{8\qz} + \sqrt{\frac{2}{\pi}} \frac{\aad}{8 \qx \qy \qz} \nonumber \\
			&+ 2 N^4 \gamma_\rho^2 (\qx^2 + \qy^2) + 2 N^4 \gamma_z^2 \qz^2 \nonumber \\
			&+ \frac{1}{24 \sqrt{2 \pi} \qz} \left[ \frac{1}{\qz^2} 
			  \RD\left( \frac{\qx^2}{\qz^2}, \frac{\qy^2}{\qz^2}, 1\right) - \frac{1}{\qx \qy} \right] \,,
	\label{eq-potential}
\end{align}
where $\RD(x,y,z) = \frac{3}{2} \int_0^\infty \left[ (x+t) (y+t)
(z+t)^3 \right]^{-1/2} \ud t$ is an elliptic integral 
in Carlson form \cite{Carlson1995}. For given physical values of the scattering
length and the trap frequencies the potential fully describes the dynamics of
the BEC in the Hilbert subspace of the variational ansatz \eqref{eq-ansatz}. In
what follows we fix the values of the mean trap frequency to
$N^2(\gamma_\rho^2\gamma_z)^{1/3} = 3.4\times 10^4$ and of the trap aspect ratio
to $\lambda=\gamma_z/\gamma_\rho = 50$, if not stated otherwise, and vary
$a/a_d$. 

Note that, because of the large aspect ratio of the trap, the dipoles are
predominantly aligned in side-by-side configuration where they repel each
other and stabilize the BEC against collapse. In the following, we will,
therefore, only consider the regime of a negative s-wave scattering length
($\aad <0$) which counteracts this effect.
\begin{figure}[t]
\centering
\includegraphics[width=\columnwidth]{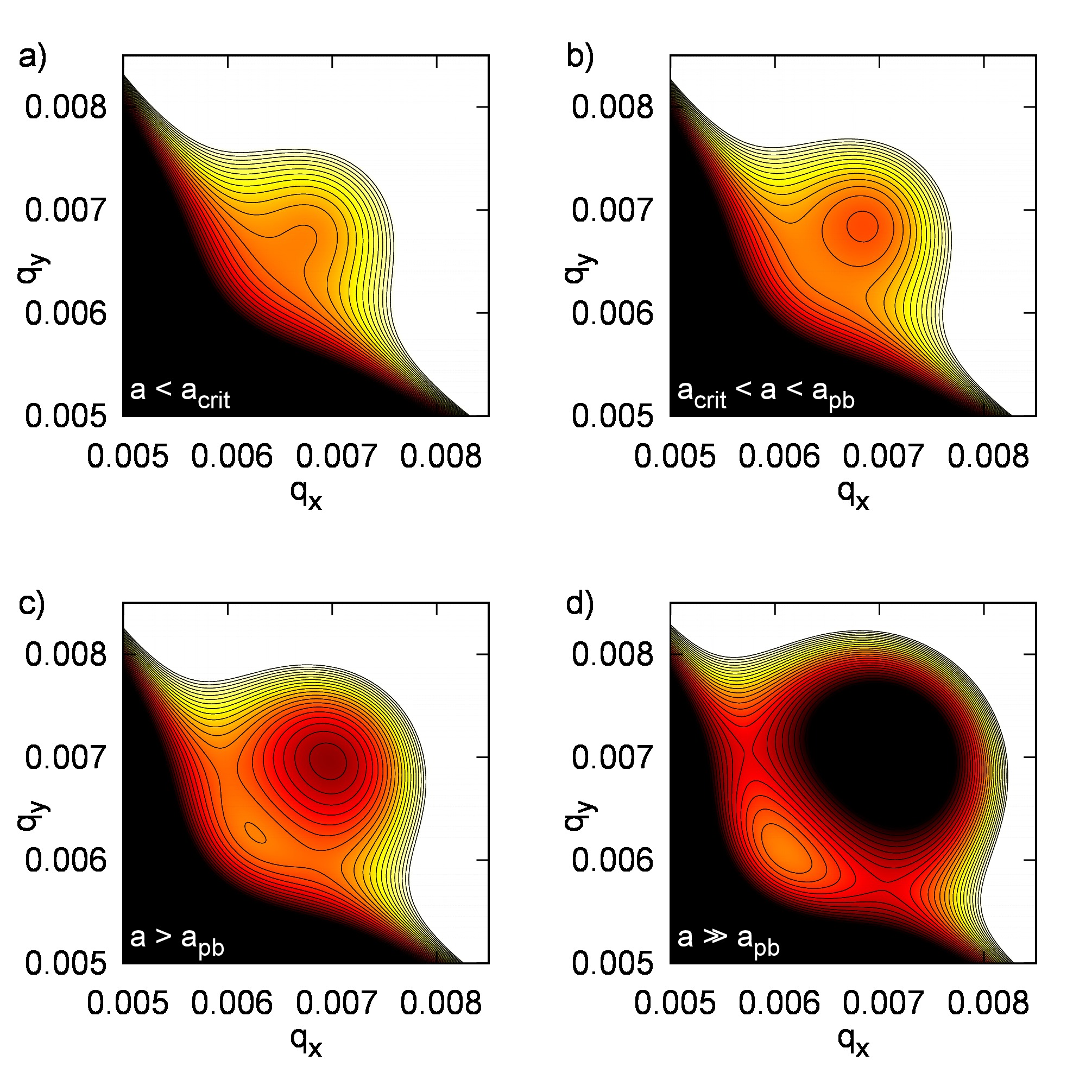}%
\caption{(Color online) Contour plots of the potential \eqref{eq-potential} in
dependence of
the generalized coordinates $\qx,\qy$ for different values of the scattering
length $\aad < \acrit$ (a), $\acrit < \aad < \apf$ (b), $\aad \gtrsim \apf$ (c)
and $\aad \gg \apf$ (d). The third coordinate $\qz = \text{const.}$ is fixed to
the value
corresponding to the cylindrically symmetric excited state. See text for
description. The color scale is different in every plot, black corresponds to
low, white to high values of $V(\qq)$.}
\label{fig-contours}
\end{figure}

Fig.\ \ref{fig-contours} shows contour plots of the potential
\eqref{eq-potential} for several values of the scattering 
length $\aad$ and fixed coordinate $\qz$. Below a critical value, $\aad <
a_\text{crit}$ (Fig.\ \ref{fig-contours}a), there exists no stationary point of
the potential. Two of these emerge in a tangent bifurcation at  $\acrit \approx
-0.22723$, and both are cylindrically symmetric. One represents the stable
ground state of the BEC, and the other is an unstable excited state (Fig.\
\ref{fig-contours}b). At a scattering length $\apf \approx -0.22657$ two
additional and
non-axisymmetric states emerge from the central saddle in a pitchfork
bifurcation, forming two satellite saddles (Fig.\ \ref{fig-contours}c--d) and
turning the central one into a rank-2 saddle. 

The potential \eqref{eq-potential} allows for a direct interpretation in terms
of reaction dynamics of thermally 
excited dipolar condensates: In the case $\acrit < \aad < \apf$, i.e.\ in the
region where only the center saddle exists (Fig.\ \ref{fig-contours}b), a
sufficient thermal excitation of the BEC may allow the system to cross the
center saddle, and to escape to $\qx,\qy \to 0$, which means the collapse of the
BEC. In this case the reaction path will always be located on the angle
bisector, and thus this represents a condensate which collapses in a
cylindrically symmetric way. The situation changes qualitatively when the
parameter region $\aad > \apf$ (Fig.\ \ref{fig-contours}c--d) is reached: Since
the
two satellite saddles are of lower energy than the central one the reaction path
now breaks the cylindrical symmetry and crosses one of the satellite saddles,
which  means that the condensate collapses with an $m=2$-symmetry. 

\subsection{Calculation of the reaction rate}
\label{sec-theory-b}

The particle number scaled reaction rate can be calculated by applying TST and
is given by \cite{Haenggi1990}
\begin{equation}
\tilde{\Gamma} = N^{2} \Gamma = \frac{1}{\tilde{\beta} Z_0}  \left( \frac{2
\pi}{\tilde{\beta}}
\right)^{\tfrac{d-1}{2}} \int_S\ud^2\vec{q}' ~ 
\ue^{- \tilde{\beta} V(\qq', q'_1=0) } \,,
\label{eq-gamma-general}
\end{equation}
where new variables $\qq'$ are defined in such a way that the reaction
coordinate $q'_1=0$ defines a dividing 
surface~$S$ that separates the configuration space into a region of reactants
(stable BEC) and products (collapsing BEC), $d$ is the system's number of
degrees of freedom, and $Z_0$ is the canonical partition function. Approximating
the potential harmonically at the ground state $(0)$ as well as at the activated
complex $(b)$ yields the reaction rate \cite{Haenggi1990}
\begin{equation}
\tilde{\Gamma} = \frac{1}{2\pi} \, \frac{\Omega^{(0)}}{\Omega_i^{(b)}} \,
\ue^{- \tilde{\beta} V_0^\ddagger} \,,
\label{eq-Gamma-conventional}
\end{equation}
where $\Omega_i^{(0)}=\prod_{i=1}^d \omega_i^{(0)}$ and
$\Omega_i^{(b)}=\prod_{i=2}^d \omega_i^{(b)}$ are the products 
of the oscillation frequencies $\omega_i^{(0,b)}$ at the ground state and the
saddle, respectively, and $V_0^\ddagger$ is the energy difference between the TS
and the ground state.

In the cases $\aad \ll \apf$ and $\aad \gg \apf$, i.e.\ far away from the
bifurcation,
Eq.\ \eqref{eq-Gamma-conventional} 
will yield an appropriate approximation for the reaction rate, since then the
reaction will either proceed over the central saddle or over one of the
satellites. (In the latter case, the rate~\eqref{eq-Gamma-conventional} must be
doubled because there are two saddles.) However, in the vicinity of the
bifurcation ($\aad \approx \apf$), Eq.\ \eqref{eq-Gamma-conventional} will fail:
Mathematically this is because one of the frequencies $\omega_i^{(b)}$ occurring
in the denominator will vanish at the bifurcation, leading to the divergence of
the reaction rate. Physically speaking, it will fail because the center and
satellite saddles are separated by energies of $\kB T$ or less, and reactive
trajectories can pass over the central saddle with nearly the same probability
as over the satellites.

Since close to the bifurcation the quadratic expansion of the potential is
obviously not adequate to reproduce the 
correct behavior, we need a more accurate approximation. It is provided by the
classifications of catastrophe theory \cite{Poston1978,Castrigiano1993}, and we
therefore apply a change of coordinates $\qq' \to \xx$ that maps the potential
$V(\qq',q_1=0)$ to a suitable normal form $V_0 + U(\xx)$. The remaining integral
in Eq.\ \eqref{eq-gamma-general} then has the form
\begin{equation}
I = \int \! \ud^2\xx \, \phi(\xx) \, \ue^{- \tilde{\beta} U(\xx)}\,,
\label{eq-definition-I}
\end{equation}
where $\phi(\xx)$ is the Jacobi determinant arising from the transformation, and the reaction rate reads
\begin{equation}
	\tilde{\Gamma} = \frac{\Omega^{(0)}}{2 \pi}\, \ue^{- \tilde{\beta}
V_0^\ddagger} \times I.
	\label{eq-Gamma-uniform}
\end{equation}

A suitable normal form describing the bifurcation of the transition state in the axisymmetric trap is
\begin{equation}
	U(\xx) = \frac{1}{4} x_2^4 + \frac{u}{2} x_2^2 + \frac{1}{2} \sum_{i=3}^d \left[\omega_i^{(b)}\right]^2 x_i^2.
	\label{eq-symmetric-cusp-normal-form}
\end{equation}
It is quadratic in all variables but one. The number and type of stationary
points of $U$ depends on the value of the 
parameter~$u$. By a suitable choice of $u$, we will reproduce the bifurcation of
saddle points that is found in the physical potential $V$.

For $x_i=0$ ($i\neq 2$) and $u<0$ the function $U(\xx)$ has a maximum at
$x_{2,\text{cs}}=0$ (center saddle) and two 
minima at $x_{2,\text{ss}} = \pm\sqrt{-u}$ (satellite saddles) with $
U(x_{2,\text{ss}}) =-u^2/4$. In the case $u>0$, $x_{2,\text{cs}}=0$ is a
minimum, and the other stationary points $x_{2,\text{ss}}$ are imaginary. With
the energy difference $\Delta V^\ddagger$ between the central saddle and the
satellite saddles we further define the unfolding parameter $u=\pm 2 \sqrt{
\Delta V^\ddagger }$ and choose it negative if all stationary states are real,
and positive otherwise. With this choice Eq.\
\eqref{eq-symmetric-cusp-normal-form} by construction reproduces the physical
energy gap of the saddle configuration over the whole range of the scattering
length $\aad$.

What remains is to determine the prefactor $\phi(\xx)$ in Eq.\
\eqref{eq-definition-I} in such a way that the flux 
integral reproduces the standard TST rate far away from the bifurcation. In the
case $u\to \infty$ (only the center saddle) we can return to the quadratic
approximation of the potential, and because the prefactor varies slowly we can
regard it as constant. In this limit we have
\begin{equation}
	I \approx \sqrt{\frac{2 \pi}{\tilde{\beta} u}} \, \phi(0)  
	\equiv \frac{1}{\Omega_\text{cs}^{(b)}}
	\label{eq-det-phi0}
\end{equation}
where ``$\equiv$'' denotes the requirement that the conventional TST result
is to be reproduced. Analogously in the limit $u \to -\infty$ we require
\begin{equation}
	I \approx 2 \times \phi(\xx_\text{ss}) \, \sqrt{\frac{ \pi}{-\tilde{\beta}
u}} \, \ue^{- \tilde{\beta} u^2/4}  
	\equiv 2\times \frac{1}{\Omega_\text{ss}^{(b)}}
	\label{eq-det-phi12}
\end{equation}
to reproduce the flux over the two satellite saddles. Since $\phi(\xx)$ must be
an even function, we finally write as 
its lowest-order Taylor expansion
\begin{equation}
	\phi(\xx) = \phi(0) - \frac{\phi(\xx_\text{ss}) - \phi(0)}{u} \, x_2^2\,, 
\end{equation}
and, once the values of $\phi(0)$ and $\phi(\xx_\text{ss})$ have been determined
from Eqs.\ \eqref{eq-det-phi0} 
and~\eqref{eq-det-phi12}, we solve the remaining integral in Eq.\
\eqref{eq-Gamma-uniform} numerically. 

In a different setting, the corrections to TST~rates that are due to rank-2
saddles were recently estimated by 
Maronsson \etal\ \cite{Maronsson2012}, who calculated the energy ridge that
connects the rank-1 saddle to the rank-2 saddles. In contrast to ours, their
method  takes account of the precise shape of the potential along the ridge. The
present approach provides a rate formula that applies on both sides of, and
arbitrarily close to, the bifurcation. It also offers the advantage of greater
simplicity because it only requires information about the saddle points
themselves. Via the frequencies $\omega_i^{(0,b)}$, the influence of degrees of
freedom transverse to the ridge is taken into account.

\section{Results}
\label{sec-results}

\begin{figure}[t!]
\centering
\includegraphics[width=\columnwidth]{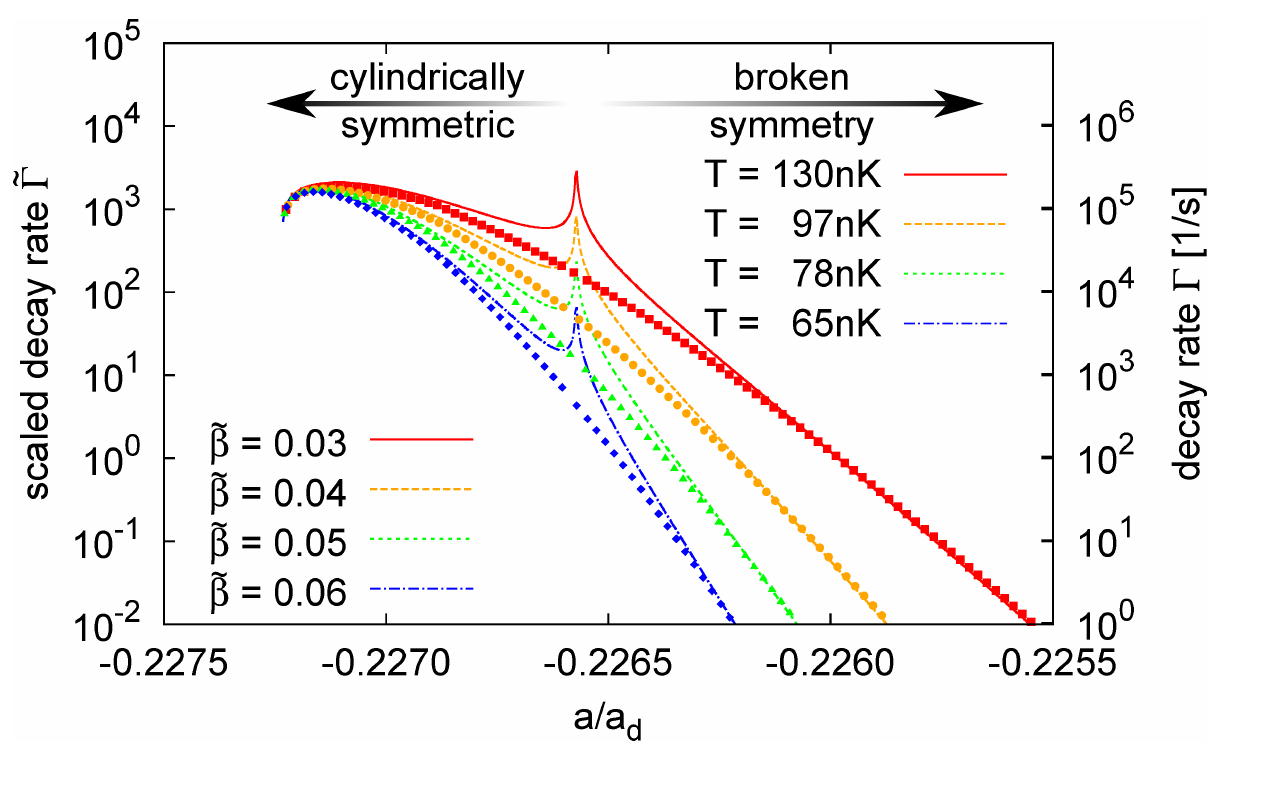}%
\caption{(Color online) Scaled decay rate $\tilde{\Gamma}= N^{2} \Gamma$ of the
dipolar BEC in
dependence of
the scattering length
$\aad$ and for different temperatures $\tilde{\beta} = N^{-1}\beta$. The lines
show the
results
calculated from the conventional TST rate formula \eqref{eq-Gamma-conventional}
and the dots show the corresponding reaction rate obtained from the uniform rate
formula \eqref{eq-Gamma-uniform}. The temperatures $T$ as well as the
decay rate $\Gamma$ have been
calculated for $^{52}$Cr BECs with a particle number of $N=50\,000$.}
\label{fig-decay-rates}
\end{figure}

Fig.\ \ref{fig-decay-rates} shows the thermal decay rates of the dipolar BEC in
leading-order TST calculated from Eq.\ 
\eqref{eq-Gamma-conventional} in comparison with the results obtained from the
uniform rate formula for the rank-2-rank-1 saddle configuration, Eq.\
\eqref{eq-Gamma-uniform}. The first case solely considers the energetically
lowest saddle(s) (lines) while the second case takes into account the complete
configuration of saddles (dots). In the calculations using the conventional TST
rate formula (lines), the divergence of the decay rate at $\aad \approx
-0.22657$ is
obvious. By contrast, the uniform solution (dots) passes the bifurcation
smoothly. We again emphasize that the collapse of the BEC will be cylindrically
symmetric on one side of the bifurcation, and symmetry-breaking on the other
side. Near the bifurcation, however, a clear distinction can no longer be made.

In the calculation the particle number scaled temperatures have been adapted
to a $^{52}$Cr BEC with a magnetic moment of $\mu = 6 \mu_\text{B}$
($\mu_\text{B}$ is the Bohr magneton) and a particle number of $N=50\,000$ as it
has
been realized experimentally by Griesmaier \etal\ \cite{Griesmaier2005}. For
this number of
bosons the values $\tilde{\beta} = 0.03$ to $\tilde{\beta} = 0.06$ correspond to
temperatures
between $T = 65\,$nK and $T = 130\,$nK which is clearly below the critical
temperature of $\Tc \approx 700\,$nK so that the treatment within the
Gross-Pitaevskii
framework is justified.

Note that, on the other hand, these temperatures are high enough to
activate collective oscillations of the BEC: In the relevant
region of the
scattering length, the frequencies of the monopole and the quadrupole mode are,
both, on the order of $\tilde{\omega} \sim 10\,000$. For the above mentioned
particle
number, this means an oscillation frequency of $\omega =
107\,\text{s}^{-1}$. Assigning to this frequency an energy of $E \sim \hbar
\omega$ as well as the temperature $T \sim E / \kB$, we find a
value of $T = 0.8\,$nK to determine the order on which collective
oscillations are activated. Thus, for the temperatures given above the latter
are sufficiently present. 

\begin{figure}[t]
\centering
\includegraphics[width=\columnwidth]{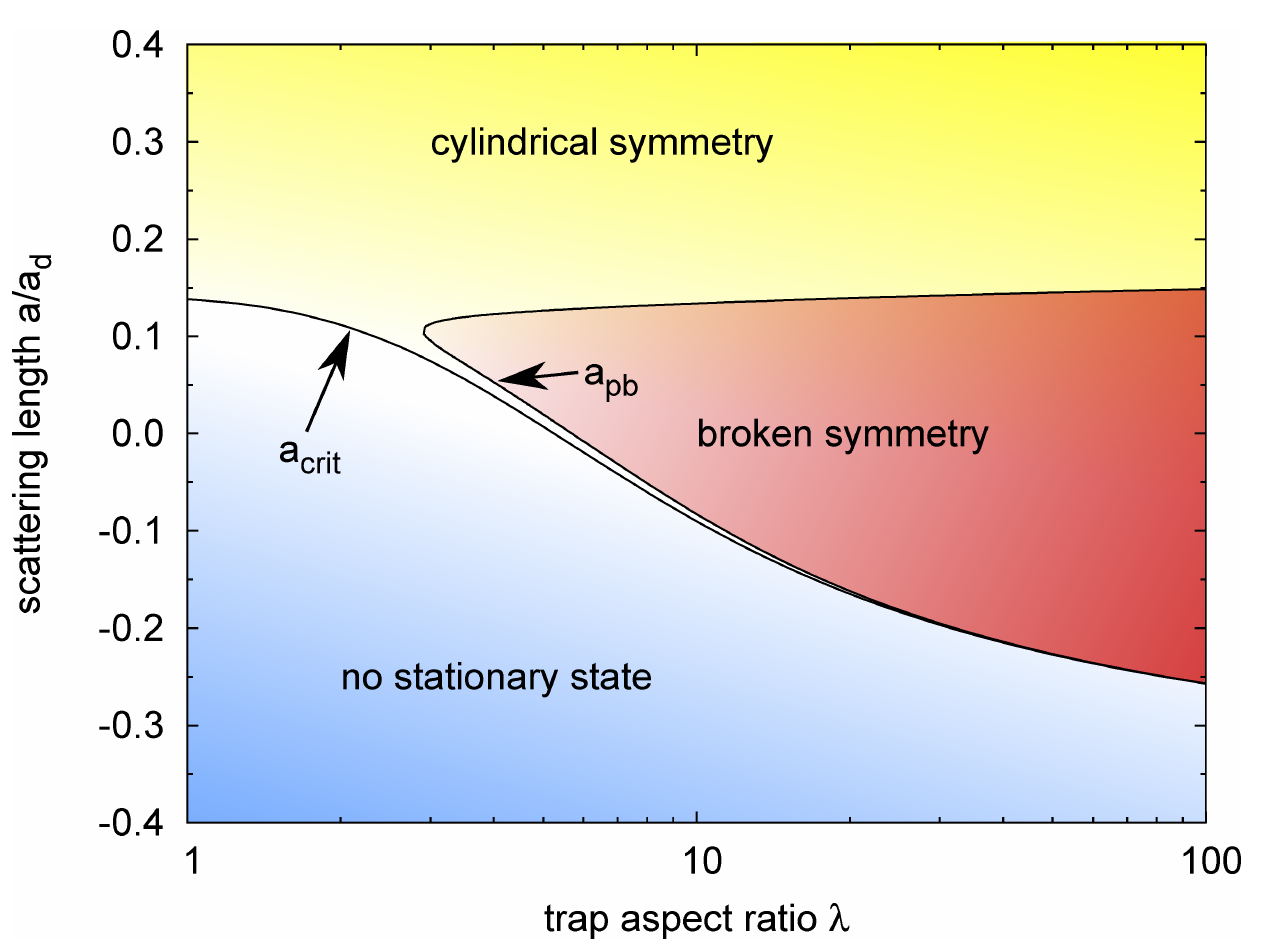}%
\caption{(Color online) Stability diagram of the unstable excited state(s) of
the GPE in
dependence of the scattering length $\aad$ and the trap aspect ratio $\lambda$.
There exists either no excited state, only the symmetric state or in addition
two symmetry-breaking states.}
\label{fig-phase-diagram}
\end{figure}

For experiments it will be of great interest in which region of the physical
parameters (trap frequency and scattering 
length) a symmetry-breaking collapse is to be expected. Fig.\
\ref{fig-phase-diagram} shows that the existence of the symmetry-breaking states
and the corresponding regions of the scattering length crucially depend on the
trap aspect ratio. While for small $\lambda \lesssim 2.8$ (including prolately
trapped condensates $\lambda < 1$, not shown) only the cylindrically symmetric
excited states exist, the additional symmetry-breaking states appear for oblate
condensates with $\lambda \gtrsim 2.8$. The more oblate the BEC the larger
becomes the region in which these states are present. In contrast, increasing
the trap aspect ratio, the parameter region of the scattering length with
$\acrit < a < \apf$ becomes smaller and vanishes for $\lambda \to \infty$. We
therefore expect the trap aspect ratio to be the decisive tool to switch between
the two scenarios in an experiment. Note that the curve in 
Fig.\ \ref{fig-phase-diagram} for the critical scattering length of course
corresponds to the one published by Koch 
\etal\ \cite{Koch2008}.

\section{Conclusion and outlook}

We have investigated a thermally excited dipolar Bose-Einstein condensate
in a cylindrically symmetric trap. Within a variational framework we
observed that the unstable excited state of the system which forms the
“activated complex” on the way to the collapse of the condensate undergoes a
bifurcation. This divides the parameter region of the s-wave scattering
length into a region with cylindrically symmetrical collapse, and one where
the collapse occurs with broken symmetry. With the help of a uniform rate
formula, we were able to calculate the corresponding reaction rate over the
whole range of the scattering length within leading order TST and to smoothly
pass the bifurcation. Moreover, the occurrence of the additional bifurcation
strongly depends on
the trap geometry which allows one to switch between the two scenarios in
experiments.

In order to improve the results quantitatively, the procedure described here can
be extended to \emph{coupled} Gaussian wave functions, which have already proven
their power to reproduce or even to exceed the quality of numerical results
\cite{Rau2010a,Rau2010b}. We have shown elsewhere
\cite{Junginger2012a,Junginger2012b} that it is possible to construct a
Hamiltonian also for the case of coupled Gaussians which then allows for the
application of TST. While in the case of a long-range $1/r$-interaction we could
show that converged results for the decay rate are only shifted to higher values
of the scattering length, the situation is different in dipolar BECs: The
bifurcation of the TS leading to the symmetry-breaking stationary states also
exists in the case of coupled wave functions, however, in the latter case even
more bifurcations occur when the number of Gaussians is increased. The even
richer thermal collapse scenarios and decay rates of 
dipolar BECs described by coupled Gaussians are a challenge for currently ongoing research.

\begin{acknowledgments}
This work was supported by Deutsche Forschungsgemeinschaft. A.\,J. is grateful for support from the Landesgraduiertenf\"orderung of the Land Baden-W\"urttemberg.
\end{acknowledgments}


%

\end{document}